\RequirePackage[2020-02-02]{latexrelease}
\documentclass[aps,prl,preprint,superscriptaddress]{revtex4}
\usepackage{graphicx}
\usepackage{dcolumn}
\usepackage{bm}
\usepackage{epstopdf}
\usepackage{csquotes}
\usepackage{amsmath}
\usepackage{color} 
\usepackage{soul} 
\usepackage{lineno}
\usepackage[]{changes} 
\newcommand{\add}{\textcolor{black}}
\newcommand{\remove}[1]{}

\begin{document}


\title{Non-Markovian Dynamics in Fiber Delay-Line Buffers }

\author{Kim Fook Lee and Prem Kumar}
\affiliation{%
Center for Photonic Communication and Computing, Department of Electrical Engineering and Computer Science, Northwestern University, 2145 Sheridan Road, Evanston, IL 60208-3112, USA}%
\email[]{kim.lee@northwestern.edu}



\date{\today}

\begin{abstract}
We study the non-Markovian effect on a two-photon polarization entangled state, in which one photon from the pair is stored in a fiber delay-line buffer. We propose a model of a photonic qubit coupled to fiber birefringence and a fiber reservoir representing the environment. We analytically derive a non-Markovian probability function for the buffered photon and its paired photon. To verify the probability function, we perform full quantum state tomography of the photon pairs. The probability function fits well with the experimental data and physical values.
Our results indicate that our quantum system \remove{is operating}\add{operates} slightly above the threshold for a non-Markovian transition. We observe a unique polarization dynamic of the buffered photon.
We further exploit \remove{the} measures of quantum mutual information \remove{for studying}\add{to study} the quantumness of the photon pairs.  \remove{We find that} Werner's well-known separability criterion occurs at a buffer time of about 0.9$\,$ms. Our results imply that \remove{the} quantum discord can surpass Werner's criterion, and hence, \remove{the} quantum bi-partite correlation can exist for \remove{a} buffer times greater than 0.9$\,$ms.

\end{abstract}


\maketitle


\section{Introduction}

Photons that traverse in optical fiber with little attenuation are ideal carriers for quantum networking. \remove{The}\add{Photonics} qubits with entanglement in polarization, time-bin, \remove{or}\add{and} other degrees of freedom are essential resources for establishing quantum communication and distributed quantum computing.
Quantum buffers made of optical fiber and free-space optics can \remove{provide}\add{operate over} the entire telecom C-band optical bandwidth and \add{have} storage time\add{s} up to \remove{$\mu\rm{sec}$}\add{microseconds} without significant deterioration of storage efficiency \cite{kflee24,pittman02,pittman23,XLi_OL_2005,Kwiat_SPIE_2021,Kwait_optica_2017,Wang_2022,Takesure_ncomm_2013,Park_OE_2019,
Clemmen_OL_2018,Mendoza_Opa_2016}. Short buffer times (a few nanoseconds) can provide active temporal source multiplexing for high-efficiency single\add{-}photon generation \cite{Park_OE_2019,Kaneda_2015,Xiong_NC2016}.
Long storage times play an important role in optical networks for packet synchronization, label processing, and contention management.
Quantum buffers based on fiber delay lines \remove{are}\add{can be} traveling buffers \cite{Zhong_2001,Kaneda_2015,Xiong_NC2016} \remove{and}\add{or} fiber loop buffers \cite{Weber_1996, kflee24}.
The traveling buffers, which can provide qubit storage time up to \remove{$\mu\rm{sec}$}\add{microseconds}, are relatively unexplored.
In general, there is a strong coupling between qubits, fiber birefringence, and a fiber reservoir,  so we need to develop a non-Markovian model to study the decoherence dynamics of qubits in a fiber-based quantum buffer.

Recent studies \cite{Giovannetti_PRL_22,Giovannetti_JPA_05,Helsen_2023} indicate that  Markovian and non-Markovian processes can be controlled \remove{for improving}\add{to improve} the quantum capacity in \remove{a} high-loss communication channels and \remove{modeling}\add{model} quantum networks. Interestingly, one can protect quantum entanglement by using the memory effect of the classical noises in the non-Markovian regime\cite{Wenjie_2023}.
The fiber-based quantum buffer is an open quantum system because the polarization degree of freedom of \add{the} photon is coupled with the frequency degree of freedom of the fiber birefringence\add{,} which is coupled with the fiber reservoir.
The density matrix of a qubit can evolve into Markovian (non-Markovian) dynamics depending on the weak (strong) coupling with the medium, respectively. The medium here is a \add{birefringent} single-mode optical fiber \remove{birefringence} and a fiber reservoir.
In a weak coupling process, the medium induces memoryless dynamics on the qubit, so \remove{that} its density matrix  $\rm{\rho (t)}$ at time t is determined by $\rm{\rho (t-dt)}$ with the infinitesimal dt. In other words, the information of a qubit should not be memory-related \remove{with}\add{to} the previous qubit. \add{In this work, we study the time evolution of a two-photon state in the medium.} The Markovian dynamic\add{s} of a qubit \remove{is}\add{are} ideal for long-distance communication networks.
The non-Markovian process, in contrast, \remove{keeps}\add{retains} the memory of the coherence dynamics of the qubit such that its density matrix $\rm{\rho (t)}$ at time t is determined by its past $\rm{\rho (t')}$  where $\rm{t'}$ is the start of the strong coupling process. It is essential to minimize the non-Markovian dynamic\add{s} of a qubit stored in a fiber-based buffer by exploring its coupling with fiber birefringence and \add{the} fiber reservoir. A comprehensive discussion of concepts, measures, and witness\add{es} of \add{the} non-Markovian effect for an open quantum system can be found in \remove{ref}\add{references} \cite{Plenio_2014,Piilo_16}. A trace distance of two quantum states \cite{Piilo_09, Piilo_10} is usually used to detect the crossover between \remove{the} Markovian and non-Markovian processes, such as the coupling between the photon polarization and its optical frequency degree of freedom in a simple titled Fabry-Perot cavity platform \cite{Guo_nphy_2011}.
The non-Markovian effect can revive entanglement after \add{the} observation of entanglement sudden death \cite{Haase_2018,Wang_2018}.

An exactly solvable master equation of an open quantum system, such as a two-level atom interacting resonantly in a reservoir formed by the Lorentzian quantized modes of a high Q cavity, is often used as an example \remove{for illustrating}\add{to illustrate} the non-Markovian dynamics of a qubit \cite{Breuer_oxford_2002,Petruccione_PRA_2006,Bellomo_PRL_2007,Chen_PRA_14,Wang_scr_17}.
Furthermore, the non-Markovian dynamics have been studied \remove{by} using quantum discord \cite{Wang_PRA_2010,Fanchini_PRA_2010}, \remove{which is} another concept of quantum correlation introduced by quantum mutual information theory. Some believe that quantum discord is more suitable for quantifying the quantum correlation of a bipartite system in non-Markovian environments. We recently used it to explore the quantum coherence of a two-qubit entangled state generated in green fluorescent protein \cite{Kflee_nc_2017,Shi_sr_2016} at room temperature. One can experimentally obtain quantum discord by calculating the total correlation and classical correlation based on the density matrix elements of a quantum state.

In this paper, one photon from the polarization entangled photon pair is stored in the buffer. We use a model to study the non-Markovianity of \remove{a photonic qubit}\add{the photon pair, where the buffered qubit state is} coupled to the fiber birefringence of a single-mode optical fiber in a fiber reservoir. We analytically derive the probability function for the buffered polarization qubit and its paired photon. Quantum state tomography (QST) is then performed on the entangled photon pairs. We obtain the probability function from the reconstructed density matrix. Our model agrees with the experimental data.\remove{The buffered photon exhibits the non-Markovianity because of the
polarization modal dispersion (PMD) and the loss in the fiber reservoir.} \add{The photon pair exhibits non-Markovianity because the buffered photon experiences the polarization modal dispersion (PMD) and loss in the fiber reservoir.}
In the experiment, we also prove the existence of the Markovian process in the scenario where we replace the buffer with a free-space attenuator.

\section{Results}
\begin{figure}[htbp]
\centering
\includegraphics[scale=0.5]{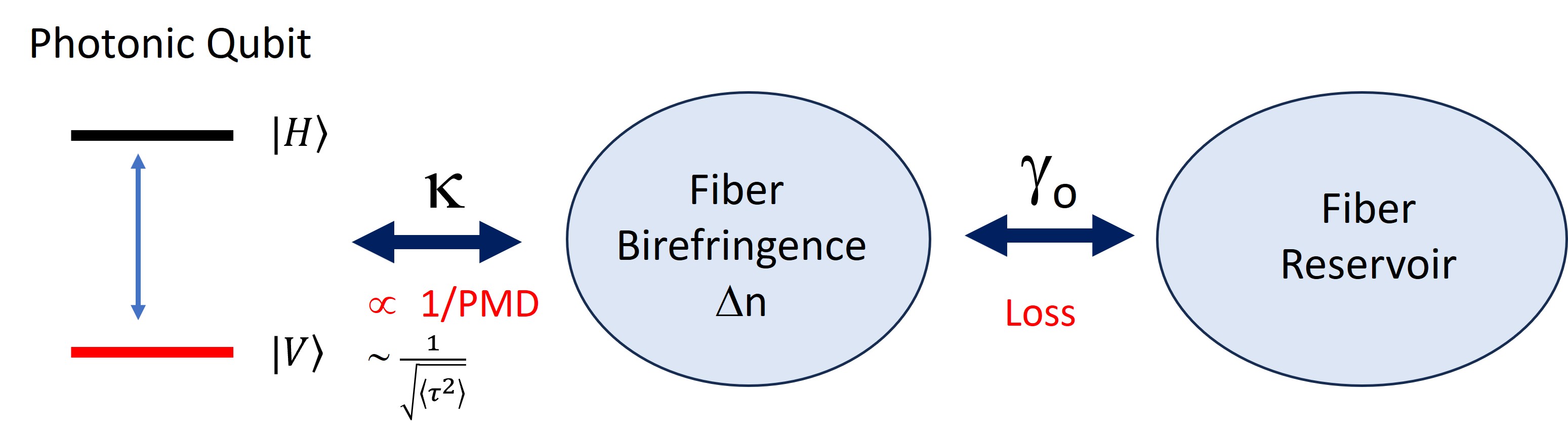}
\caption{\textbf{A simple model consists of a photonic qubit, fiber birefringence in optical fiber, and a fiber reservoir representing the environment\add{al} decoherence. The $\kappa$ is the coupling constant between the qubit and fiber birefringence ($\Delta n$). The $\kappa(\rm{sec}^{-1})$ is inversely proportional to PMD ($\sim \frac{1}{\sqrt{\langle \tau^{2} \rangle}}$), where $\langle \tau^{2} \rangle$ is \add{the} differential group delay (DGD). The fiber birefringence is then coupled with a fiber reservoir with the coupling constant $\gamma_{\circ}(\rm{sec}^{-1})$, which is responsible for the loss of qubit information. \remove{such as decoherence.}}}
\label{Figure1}
\end{figure}

Here, we consider a qubit-fiber-reservoir model consisting of a photonic qubit, a single-mode optical fiber, and a fiber reservoir representing the environment as shown in Fig.1. (see Methods for the atom-cavity-reservoir model).
When a photon propagates in a long optical fiber, its polarization state and transmission rate will be degraded by the fiber birefringence and a fiber reservoir. The fiber reservoir is the medium where a polarization qubit interacts \add{and} exchanges energy with the environment. The polarization of the photon with \add{a} certain optical frequency bandwidth is coupled with the fiber birefringence, which is coupled with the fiber reservoir. In other words, the PMD effect is coupled to the environment via the fiber reservoir. The photon polarization state will experience \remove{dephasing}\add{depolarization} caused by the PMD and the absorption/scattering process in the fiber. The absorption and re-emission of a photon state in the fiber will decohere a quantum state.
A photonic qubit with polarization state H or V is coupled with the fiber birefringence by a coupling constant $\kappa \propto \frac{1}{\rm{PMD}}$. A photonic qubit in a single optical frequency mode will couple weakly with the fiber birefringence, \add{experiencing less dispersion (large $\kappa$)}, and hence \remove{experiencing} less depolarization effect.
If we increase the frequency bandwidth of the photonic qubit in a steep dispersion region, the $\kappa$ becomes small. The qubit will start to experience the \remove{PDM}\add{PMD} effect, which is given by the square root of time variance \remove{$\langle \tau^{2} \rangle$}\add{$\sqrt{\langle \tau^{2} \rangle}$} of differential group delay (DGD) \cite{Kogelnik_PMD_2002,Wai_jlt_1996,Brodsky_jlt_2006}\remove{ or $\sqrt{\langle \tau^{2} \rangle}$}. Therefore, the strength of the  PMD effect is inversely proportional to the $\kappa$. In other words, one can enhance the PMD effect on the photonic qubit by increasing \remove{their}\add{its} differential group delay\remove{s}, such as \add{by} inserting a few polarization-maintaining fiber jumpers in \remove{their}\add{its} channel\remove{s}.
The same photon with optical frequency $\omega$  can be spontaneously absorbed and scattered at a rate of $\gamma_{\circ}$ due to the coupling between the fiber birefringence and the reservoir.
The $\frac{\gamma_{\circ}}{c} \propto\mu$ is the loss parameter per unit length of the photon due to photon absorption/scattering in the fiber reservoir.
(see Methods for the probability function of the photon state and how the relative strength of the $\gamma_{\circ}$ and $\kappa$ can determine the photon state undergoing Markovian and non-Markovian processes).
Two factors contribute to the loss parameter. The first is the propagation loss of the photonic qubit in the fiber. The second is the loss of information on the photonic qubit, which is its polarization property. The \remove{environment}\add{environmental} depolarization process can degrade the polarization extinction ratio of the qubit. The propagation loss of the photonic qubit can cause an increase in \remove{the} accidental coincidence and reduce the fidelity of the quantum state. In this paper, we subtract the accidental coincidence of the signal and idler so that we can isolate the contribution of the propagation loss to the reconstructed density matrix and study the polarization property of the buffered qubit, which is the idler photon.

\begin{figure}[htbp]
\centering
\includegraphics[scale=0.5]{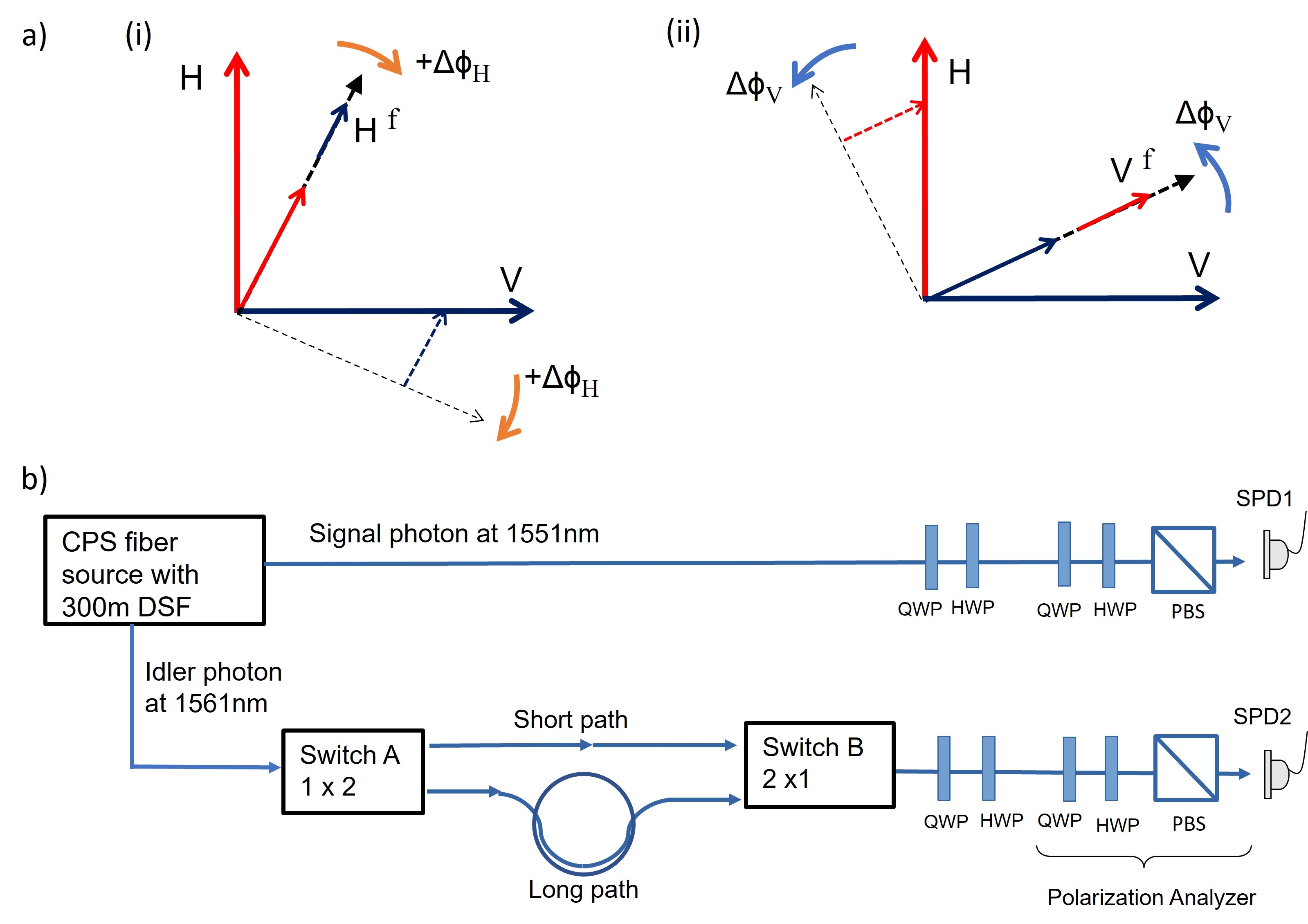}
\caption{\textbf{a) H(V) analog to the excited state (ground state) of a two-level atomic system. The \add{dynamic process} \remove{geometry picture} for the change of \add{the idler's} polarization state (i) $\rm{H} \rightarrow \rm{H}^{f}$ with $\Delta \phi_h = \Delta \phi$ and (ii) $\rm{V} \rightarrow \rm{V}^{f}$ with $\Delta \phi_v = - \Delta \phi$ . Small arrows along the  $\rm{H}^{f}$ and $\rm{V}^{f}$ show the contribution from the H and V. The phase fluctuation $\Delta\phi$ due to PMD degrades the polarization extinction ratio of the idler. b) Experiment setup including the buffer made of two low-loss switches with a short path and a long path. We use different fiber lengths in the long path to change the buffer time. \add{Polarization analyzer (PA)
consists of a quarter-wave plate (QWP), half-wave plate (HWP), and polarization beam splitter (PBS).}}}
\label{Figure2}
\end{figure}

In our two-qubit system, the photon pair is initially prepared in the entangled state $|\psi_{\circ}\rangle=\frac{1}{\sqrt{2}}[|H_s H_i\rangle + |V_s V_i\rangle]$.
One photon from the pair is sent to a buffer with \remove{the}\add{a} fiber length of L. Let's say the photon of interest is the idler, which is experiencing \add{both}\remove{the} PMD and \remove{the} loss in the buffer. The initial state of the idler (signal) is $|H_i\rangle +|V_i\rangle$ ($|H_s\rangle +|V_s\rangle$), respectively. Due to \remove{depolarizing}\add{depolarization} caused by the PMD and fiber reservoir, the final components $H^{f}_i$ and $V^{f}_i$ can be written in terms of  $H_i$ and $V_i$ as

\begin{eqnarray}
H^{f}_i = H_i \rm{cos(\Delta\phi_h)} + V_i \rm{sin( \Delta\phi_h)}\nonumber\\
V^{f}_i = V_i  \rm{cos(\Delta\phi_v)} - H_i \rm{sin( \Delta\phi_v)}
\label{eq:01}
\end{eqnarray}
where $\Delta\phi_{h(v)}$ is the amount of phase change caused by the PMD effect.
The phase change $\Delta\phi_{h(v)}=\Delta\omega\,\sqrt{\langle \tau^{2} \rangle}$  is related to the square root of time variance \remove{$\langle \tau^{2} \rangle$}\add{$\sqrt{\langle \tau^{2} \rangle}$} of differential group delay (DGD) \cite{Kogelnik_PMD_2002,Wai_jlt_1996,Brodsky_jlt_2006} or $\sqrt{\langle \tau^{2} \rangle} \propto L^{1/2}$, where L is the length of the fiber buffer.
The $h$ ($v$) sign denotes the phase change associated with the polarization state $H_i$ ($V_i$). \add{The dynamic process of the idler's polarization state described by Eq.(1) is geometrically illustrated in Fig.2(a).}
When \remove{the} $\Delta \phi_h$  and $\Delta \phi_v$ are  zero, \remove{the} $H^{f}_{i}\rightarrow H_{i}$ and $V^{f}_{i} \rightarrow V_{i}$. The polarization states of the idler do not \remove{change because of}\add{depolarize due to}  the absence of the PMD effect.   When \remove{the} $\Delta \phi_h$ and $\Delta \phi_v$ are $+90^{\circ}(-90^{\circ})$, $H^{f}_{i} \rightarrow V_{i} (-V_{i})$ and $V^{f}_{i} \rightarrow -H_{i}(+H_{i})$, respectively. The polarization states of the idler are completely flipped.
\add{When $\Delta \phi_h = \Delta \phi$, the final component $H^{f}$ of the idler consists of two components: the $\rm{cos(\Delta\phi)}$ of the initial component $H_{i}$ and the $\rm{sin(\Delta\phi)}$ of the initial component $V_{i}$, as shown in Fig.2(a)(i). The $H$ component of the idler's polarization state is decreased by an amount of $\rm{cos(\Delta\phi)}$. The polarization extinction ratio of the idler photon is degraded because of the creation of $V$ component due to the PMD effect. A similar process occurs for $\Delta \phi_v = - \Delta \phi$; the final  component $V^{f}$ of the idler consists of the $\rm{cos(\Delta\phi)}$ of the initial component $V_{i}$ and the $\rm{sin(\Delta\phi)}$ of the initial component $H_{i}$, as shown in Fig.2(a)(ii).}
\remove{The geometrical meaning for these notations is shown in Fig.2(a).}
The phase fluctuation of $\rm{cos(\Delta\phi)}$ and $\rm{sin(\Delta\phi)}$ degrades the polarization extinction ratio of the idler.
One can interpret that the final idler polarization state $H^{f}_i$ is related to its initial states $H_i$ and $V_i$ in the past \remove{because of}\add{due to} the non-Markovian effect caused by the PMD.
We assume that the signal is propagating in a short fiber length before it is detected.
We then obtain the probability function $\mathcal{P}_f =|\langle \psi_{\circ} | \psi^{f}\rangle|^2$ as,
\begin{eqnarray}
\mathcal{P}_f = |\mathcal{A} [\rm{cos(\Delta\phi_h)}+\rm{sin(\Delta\phi_h)}-\rm{sin(\Delta\phi_v)}+\rm{cos(\Delta\phi_v)}]|^{2}
\label{eq:02}
\end{eqnarray}
where $\mathcal{A} \mathcal{A}^{*} = e^{-2\mu L} $ is introduced due to the loss parameter in the fiber reservoir on the polarization properties experienced by the surviving idler photons. In our experiment, we diminish the contribution of the propagation loss to the probability function by subtracting \add{out} the accidental coincidence \add{counts} so that we can measure the probability function, which is degraded by \remove{the}  $\mathcal{A} \mathcal{A}^{*} = e^{-2\mu L} $ due to the fiber reservoir.

We explore \add{the} non-Markovian dynamics of the entangled state by simplifying Eq.(2) into two cases: the magnitude of $\Delta\phi_h=\Delta\phi_v$ and $\Delta\phi_h\neq \Delta\phi_v$.
For the case $\Delta\phi_h = \Delta\phi_v = \Delta\phi$, we have the first scenario where the idler polarization states $H_i$ and $V_i$ rotate in opposite directions, i.e, we substitute the terms $+\Delta\phi_h$ and $-\Delta\phi_v$ or $-\Delta\phi_h$ and $+\Delta\phi_v$ in Eq.(2). The $+$ ($-$) value of the $\Delta\phi_{h(v)}$  corresponds to the rotation of the polarization state in the clockwise (anti-clockwise) direction. The probability function $\mathcal{P}_f$ is simplified as
\begin{eqnarray}
\mathcal{P}_a \propto e^{-2\mu L} [\rm{cos(\Delta\phi_{1})} \pm \rm{sin(\Delta\phi_{1})}]^{2}
\label{eq:03}
\end{eqnarray}
where $\Delta\phi_{1}= \Delta\omega\,\sqrt{\langle \tau_{1}^{2} \rangle}$.
Eq.(3) \remove{has a similar form as}\add{is similar in structure to} Eq.(8) in Methods for a two-level atom interacting in a high Q-cavity with a memoryless reservoir. This implies that the idler is experiencing the non-Markovian process due to the coupling between the PMD effect and the loss in the fiber reservoir.
Let's consider the second scenario where the idler polarization states $H^{f}_i$ and $V^{f}_i$ rotate in the same directions, i.e., we substitute the terms $+\Delta\phi_h$ and $+\Delta\phi_v$ or $-\Delta\phi_h$ and $-\Delta\phi_v$ in Eq.(2). Then the probability function $\mathcal{P}_f$ becomes:
\begin{eqnarray}
\mathcal{P}_{sy} \propto e^{-2\mu L} [\rm{cos(\Delta\phi_{2})}]^{2}
\label{eq:04}
\end{eqnarray}
where $\Delta\phi_{2}= \Delta\omega\,\sqrt{\langle \tau_{2}^{2} \rangle}$. \remove{The} Eq.(4) \remove{has the same form as}\add{has a similar form to}  Eq.(9) in the Methods section.
For the case $\Delta\phi_h \neq \Delta\phi_v$, the complete probability function is shown in detail in the Methods section.

Later, we will verify the probability functions of Eq.(3) and Eq.(4) and also Eq.(8) and Eq.(9) in the Methods by launching one photon from the entangled photon pair in a quantum buffer made of different fiber lengths, as shown in Fig.2(b).
We use a counter-propagating scheme (CPS) \cite{Kflee_ol_2006,Sua_ol_2014} to generate \add{the} two-photon polarization-entangled state $|\psi_{\circ}\rangle$ through a four-wave mixing process in \remove{a} 300$\,$ \add{meters} of dispersion-shifted fiber. The detailed experimental setup is shown in Fig.1S in the Supplementary.  The signal and idler photons are separated by the DWDMs (dense wavelength-division \remove{multiplexing}\add{multiplexers}). The signal photon is sent to the polarization analyzer ($\rm{PA_{s}}$). The idler is sent to the buffer, which consists of two low-loss switches (insertion loss $\approx$ 0.01dB) and a single-mode optical fiber, as shown in Fig.2(b). We can select the idler to pass a long or short fiber length.  When the idler is selected to pass the long fiber, \remove{then the idler}it is buffered \remove{with}\add{for} the duration of $\frac{L}{c/n_{r}}$, where $L$ is the length of the fiber and $n_{r}$ is the refractive index of the fiber. The idler is then directed to the polarization analyzer ($\rm{PA_{i}}$). \add{We use a polarized classical light source to align the optical components in the polarization analyzers. The first pair of QWP (quarter-wave plate) and HWP (half-wave plate) is used to align the light polarization (H or V) along the principal axes of the second pair of QWP and HWP, and the polarizer}
\remove{The first pair of the quarter-wave plate (QWP) and half-wave plate (HWP) is used to compensate for the birefringence of the fiber.} The second pair of QWP and HWP is used to implement quantum state tomography.
The timing of detecting the buffered photon can cause a non-Markovian effect on the photon state \cite{Helsen_2023,Humprey_Nature_18}.
In this buffer configuration, the qubit storing process is initialized by two units of  1$\times$2 switch. This switching process can impose loss and decoherence, such as the amplitude or phase damping process.  We use two units of 1$\times$2 switch instead of one unit of 2$\times$2 switch because the insertion loss of 1$\times$2 switch \add{(0.01$\,$dB) is much lower than that of the} 2$\times$2 switch (1.2$\,$dB). To avoid decoherence due to timing issues on the photon detection, we operate the switch \remove{with}\add{at} low speed, i.e., 1$\,$Hz. The slow switching rate allows us to take data when the switches are both \add{fully} turned on. \remove{So}\add{Thus}, we can exclude the data \add{that} appears in the rising and falling edges of the switches. In other words, we diminish the non-Markovian effect \remove{which}\add{that} can be caused by the switching process \cite{kflee24}.
We use an interface program to turn \add{the two switches} on and off \remove{the two switches} synchronously at the rate of 1$\,$Hz. The data is taken \remove{on}\add{during} the duration when both switches are completely turned on or off.
We then perform quantum state tomography (QST) \remove{by} using two single-photon detectors (NuCrypt CPDS-4) operated at 50$\,$MHz.

We characterize the prepared state $|\psi_{\circ}\rangle = (|H_{s} H_{i}\rangle + |V_{s}V_{i}\rangle)/\sqrt{2}$ \remove{by} using the standard method of QST \cite{James_pra_2001,White_pra_2001,Munro_pra_2001}. There are a total of 16 settings for the HWPs and QWPs in the $\rm{PA_{s}}$ and $\rm{PA_{i}}$.
For each setting, we measure the coincidence counts (CC) and accidentals (AC) with an integration time of 2 seconds or 100M sampling gates \remove{when during}\add{while} the switches are turned on.
Using \remove{the} maximum likelihood estimation, we reconstruct the $4 \times 4$ density matrix of the final state $\rho^{f}$ in the HV basis.
We construct the density matrix $\rho^{f}$ based on the true coincidence counts, i.e., the accidental coincidence \add{counts} are subtracted. In other words, \add{noise photons generated via Raman scattering} \remove{the Raman photon} and loss-induced accidentals do not contribute to this study.
We repeat the state tomography for different fiber lengths from 0.0$\,$m up to 25$\,$km. The 0.0$\,$m fiber length corresponds to the short path in the buffer.

We use \remove{the} quantum discord to characterize the non-Markovian dynamics of the buffered photon \add{and its paired photon}.
It is of both fundamental and practical interest to differentiate classical and quantum correlations within the total correlation for the buffered photon experiencing a non-Markovian environment.
Two scenarios can contribute to the classical information: (1) \add{the} environment projects or makes a measurement on the idler qubit state and creates the classical state, (2) the environment's perturbation can dechore the information of the idler qubit state such as degradation of the polarization properties. Scenario (1) is the classical correlation \remove{which}\add{that} will appear at the coincidence and accidental coincidence. We can avoid this classical correlation by subtracting \add{out} the accidental \add{coincidence coounts}. Scenario (2) can create a mixture of product states. We study the quantum discord under scenario (2), which \remove{can} allow us to have a buffer time $>$ 0.9$\,$ms. For this purpose, we can accurately treat the final state as the Werner state.
The Werner state is the mixture of the maximally entangled state with the probability $P$ and the maximally mixed state, as given by \cite{Werner_pra_1989,Munro_pra_2001},
\begin{eqnarray}
\mathcal{W} = P|\psi_{\circ} \rangle \langle \psi_{\circ}| + \frac{1-P}{4} I.
\label{eq:05}
\end{eqnarray}
where the $I$ is the identity matrix.  We calculate the average \remove{of the} probability $P=\sum^{n=6}_{i} p'_{i}/n $, where $p'_{1(2)} = 4 \rho'_{11(44)}-1$, $p'_{3(4)} = 2 \rho'_{14(41)}$, and $p'_{5(6)}=1-4\rho'_{22(33)}$. The density matrix elements $\rho'_{11}$, $\rho'_{44}$, $\rho'_{14}$, $\rho'_{41}$, $\rho'_{22}$, and $\rho'_{33}$ are obtained from the final state.
The measured average probability $P$ is the non-Markovian probability function p(t). ( See Methods for amplitude damping on one photon from the two-photon Werner state).

We plot the average probability $P$ of the Werner state as a function of buffer time $t$, as shown (red solid circle) in Fig.3.
We fit the experimental data with the summation of the probability functions $\mathcal{P}_{a}$ of Eq.(3) and $\mathcal{P}_{sy}$ of Eq.(4).
With the substitution of $L = (c/n_{r}) t$ and $\sqrt{\langle \tau^{2} \rangle}=D_{p}\,\sqrt{L}$, the $\Delta\phi_{1,2}$ in the harmonic functions of $\mathcal{P}_{a}$ and $\mathcal{P}_{sy} $ \remove{is}\add{are} then given by \remove{a} $\Delta\phi_{1,2}= \Delta\omega\,\sqrt{\langle \tau_{1,2}^{2} \rangle} = \Delta\omega\, D_{p1,p2}\, \sqrt{\frac{c\cdot t}{n_{r}}}$, respectively, where $D_{p1,p2}$ is the PMD parameter in unit\add{s} of $\rm{picoseconds}/\sqrt{km}$. Let's denote $\mathcal{P}_{a,sy}=\mathcal{P}_{a} + \mathcal{P}_{sy}$.
The good fit of $\mathcal{P}_{a,sy}$ exhibits a small modulation on the exponential function of $e^{-2\mu\frac{c}{n_{r}} t}$, as shown (red line) in Fig.3. The modulation is caused by the harmonic function of $\sqrt{t}$ in $\mathcal{P}_{a,sy}$. This implies that the probability $P$ of the Werner state exhibits non-Markovian dynamics. We substitute the experiment\add{al} value of $\Delta\omega = 2\pi\times200\rm{GHz}$ in $\mathcal{P}_{a,sy}$. We obtain $D_{p1}=0.0017\,\frac{ps}{\sqrt{km}}$ and $D_{p2}=0.047\,\frac{ps}{\sqrt{km}}$.
\add{The  $D_{p2}\,> \,D_{p1}$ indicates that the harmonic function of Eq.(4) is more dominant than Eq.(3).}
Then, the total PMD parameter $D_{p} = D_{p1}+D_{p2} = 0.05\,\frac{ps}{\sqrt{km}} $, which is in agreement with the PMD link value of $\leq 0.1 \frac{ps}{\sqrt{km}}$ (Corning SMF-28 Optical Fiber data sheet).
We also obtain the loss parameter $\mu=6.0 \times 10^{-6}\, m^{-1}$ from the fitting of the experimental data. We can rewrite the loss parameter $\mu $ as 0.006/km or 0.025$\,$dB/km.  This value is below 0.05$\,$dB/km, which is the induced attenuation caused by \remove{the} environmental effects such as temperature and humidity \add{(Corning SMF-28 Optical Fiber data sheet)}.

We also fit the experimental data with the summation of the probability functions $p_{1}(t)$ of Eq.(8) and $p_{2}(t)$ of Eq.(9) in \add{the} Methods \add{section}.
Let's denote $p_{3}(t)=p_{1}(t)+p_{2}(t)$.
The $p_{3}(t)$ exhibits stronger oscillation than $\mathcal{P}_{a,sy}$ within the error bar of the data, as shown (blue line) in Fig.3. The equation of $p_{3}(t)$ indicates stronger non-Markovian dynamics than the $\mathcal{P}_{a,sy}$ because the harmonic function of $p_{3}(t)$ is a function of $t$.
We obtain the parameters $\kappa_{1}=753\,{s}^{-1}$, $\kappa_{2}=3528\,{s}^{-1}$, and $\gamma_{\circ}=16292\,{s}^{-1}$ from the fitting of the experimental data.
\add{The $\kappa_{2}\,>\,\kappa_{1}$ indicates that the harmonic function of Eq.(9) is more dominant than Eq.(8).}
The total $\kappa= 4281\,{s}^{-1}$.  We have $4\times \kappa = 17124\,{s}^{-1}$, \add{which is} greater than $\gamma_{\circ}$, satisfying the non-Markovian criterion, i.e., $4 \kappa  \geq \gamma_{\circ}$. Our experimental data confirms the non-Markovian dynamics of the buffered qubit \add{and its paired photon}.

We further prove the existence of the Markovian process if the idler photon is not buffered.
We replace the buffer in the idler channel with a free-space attenuator and adjust the attenuation $\alpha $ such that it equals the propagation loss (0.2$\,$dB/km) $\times$ the fiber length. We then perform the quantum state tomography on the signal and idler. We obtain the average probability $P$ \remove{by} using the same approach as discussed above. We plot the probability $P$ (blue data point) and also fit the data with \remove{any}\add{an} arbitrary exponential function as shown (blue dotted line) as shown in Fig.3. This indicates that the idler in \add{the} free-space configuration is experiencing \remove{the}\add{a} Markovian \remove{dynamics}\add{environment} without the fiber buffer, i.e., without the coupling to the fiber birefringence and the fiber reservoir.

\begin{figure}[htbp]
\centering
\includegraphics[scale=0.7]{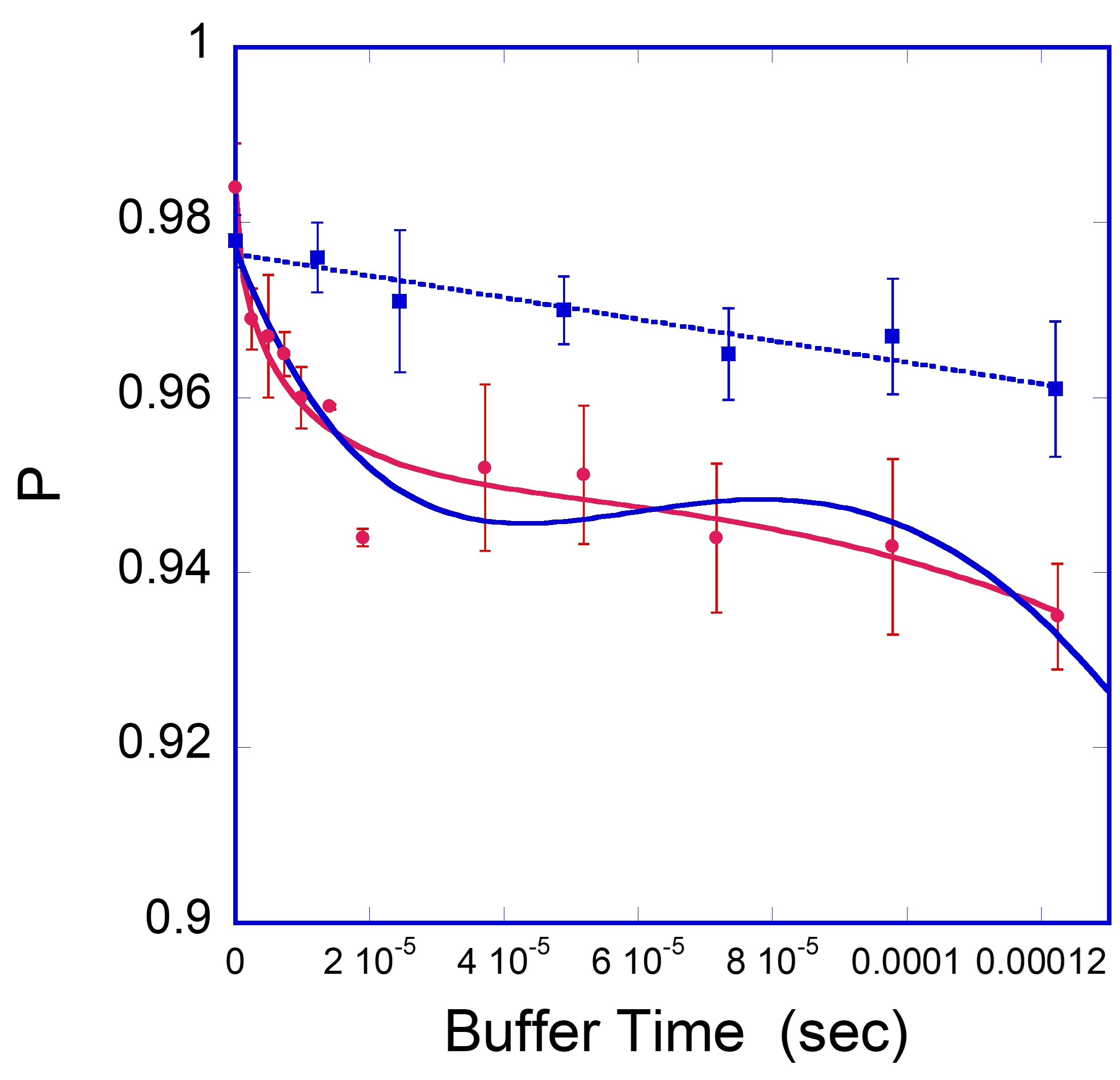}
\caption{\textbf{The measured probability $P$ of the Werner state as a function of buffer time for the buffer experiment (red solid circle) and the free-space attenuator (blue solid square). We convert the amount of the free-space attenuation ($\times$ 1/(0.2dB/km)) to buffer length and then to buffer time. The blue line (red solid line) is the fitting curve of the equation $p_{3}(t)$ (the function $\mathcal{P}_{a,sy}$), respectively. The blue dotted line is the fitting curve of an exponential function for the free-space attenuation.}}
\label{Figure3}
\end{figure}

Since the probability functions $\mathcal{P}_{a,sy}$ and $p_{3}(t)$  fit well with the measured probability $P$ of the Werner state, we can use \remove{the} $\mathcal{P}_{a,sy}$ and $p_{3}(t)$ for quantum mutual information to quantify the quantum and classical correlation of the buffered photon and its paired photon. For a bipartite system, the total correlation, which is a measure of quantum mutual information, is given by $\mathcal{I}(\rho^{f}) = \mathcal{C}(\rho^{f}) + \mathcal{Q}(\rho^{f}) $, where \remove{the} $\mathcal{C}(\rho^{f})$ is \add{the} classical correlation and \remove{the} $\mathcal{Q}(\rho^{f})$ is \add{the} quantum discord. For the Werner state, the quantum discord is given by,
\begin{eqnarray}
\mathcal{Q}(\rho^{f})=\mathcal{I}(\rho^{f})-\mathcal{C}(\rho^{f})
\label{eq:06}
\end{eqnarray}
where the total correlation and the classical correlation are given by \cite{Luo_pra_2008,Donnert_nmeth_2007},
\begin{eqnarray}
\mathcal{I}(\rho^{f})=\frac{3(1- P)}{4}\rm{log}_{2}(1- P)+\frac{(1+3 P)}{4}\rm{log}_{2}(1+3 P)\nonumber\\
\mathcal{C}(\rho^{f})=\frac{(1- P)}{2}\rm{log}_{2}(1- P)+\frac{(1 + P)}{2}\rm{log}_{2}(1 + P).
\label{eq:07}
\end{eqnarray}
We also obtain the concurrence $\mathcal{C}_{n}(\rho^{f})=\rm{max}\{0, \frac{(3 P-1)}{2} \}$ for the Werner state in the system.
\begin{figure}
\centering
\includegraphics[scale=0.5]{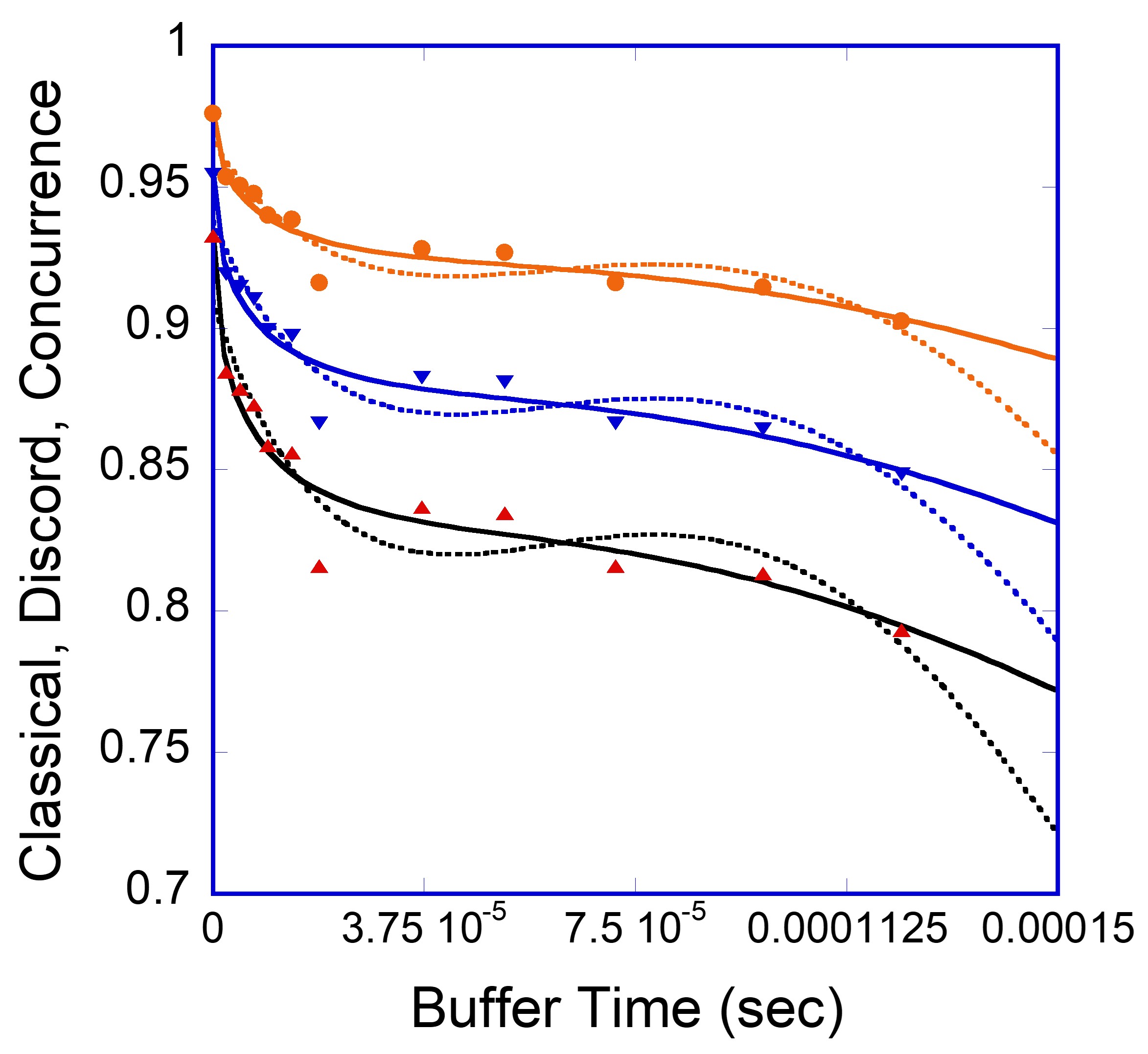}
\caption{\textbf{The experimental data: quantum discord (blue triangle), classical (red triangle), and concurrence (orange solid circle). The quantum discord $\mathcal{Q}$ (blue line and blue dotted line), classical $\mathcal{C}$ (black line and black dotted line), and concurrence $\mathcal{C}_{n}$ (orange line and orange dotted line) of the Werner state  as a function of buffer time by using the probability function $\mathcal{P}_{a,sy}$ and $p_{3}(t)$, respectively. }}
\label{Figure4}
\end{figure}
We plot the quantum discord $\mathcal{Q}(\rho^{f})$, classical correlation $\mathcal{C}(\rho^{f})$, and concurrence $\mathcal{C}_{n}$ as a function of the buffer time, as shown in Fig.4.
The experimental data is calculated based on \remove{the} Eq.(6) where the probability $P$ is obtained from the measured density matrix of the Werner state. We also calculate the quantum discord, classical, and concurrence \remove{by} using the $P=\mathcal{P}_{a,sy}$ (solid line) and the $p_{3}(t)$ (dotted line) as shown in Fig.4. See Fig.S2 in the Supplementary for the extrapolation plot  of the $\mathcal{I}(\rho^{f})$, $\mathcal{C}(\rho^{f})$, $\mathcal{Q}(\rho^{f})$ and concurrence $\mathcal{C}_{n}$ as a function of the buffer time up to 1.5$\,$ms.
When $ P > \frac{1}{3}$, which is the well-known Werner's criterion for \remove{inseparable}\add{inseparability}, the \remove{concurrent}\add{concurrence} is not zero,  indicating that the Werner state is still entangled or inseparable\cite{Zurek_prl_2001}. When the $ P = \frac{1}{3}$, the \remove{concurrent}\add{concurrence} is 0.  This occurs at the buffer time \remove{=}\add{of} 0.9$\,$ms (190$\,$km), which is numerically obtained from the fitting curve of the function $\mathcal{P}_{a,sy}=\frac{1}{3}$ (see Fig.S2 and Fig.S3(a) in the Supplementary). As for the function $p_{3}(t)=\frac{1}{3}$, this \remove{falls}\add{occurs} at the buffer time \remove{=}\add{of} 0.4$\,$ms (80$\,$km) (see Fig.S2 and Fig.S3(b) in the Supplementary). The strength of the non-Markovian effect can determine the buffer time (fiber length) where the buffered photon \add{and its paired photon} can still preserve quantum entanglement. Since the function $\mathcal{P}_{a,sy}$ is a better fit, our data indicates that the fiber-based quantum buffer can preserve quantum entanglement with a buffer time of up to 0.9$\,$ms.

Another interesting feature portrayed by the Werner state is the relationship between its concurrence, quantum discord, and classical correlation. For the function $\mathcal{P}_{a,sy}$ in the range $0.523 < \mathcal{P}_{a,sy} < 1$, the concurrence is larger than the quantum discord and classical correlation for the buffer time (length) from 0.0$\,$s(0.0$\,$m) to 0.7$\,$ms(145$\,$km). The $\mathcal{P}_{a,sy}=\add{0.523}$\remove{0.532} occurs at the buffer time \remove{=}\add{of} 0.7$\,$ms, as shown in Fig.S2 and Fig.S3(a) in the Supplementary. Similarly, for the function $p_{3}(t)=\add{0.523}$\remove{0.532}, this \remove{falls}\add{occurs} at the buffer time \remove{=}\add{of} 0.3$\,$ms (64$\,$km), as shown in Fig.S2 and Fig.S3(b) in the Supplementary. The most important feature of the quantum mutual information is \add{that} the quantum discord $\mathcal{Q}(\rho^{f})$ \add{is greater than} the concurrence and the classical correlation. This occurs at the buffer time \remove{larger}\add{greater} than 0.7$\,$ms(0.3$\,$ms) for the function $\mathcal{P}_{a,sy}$ ($p_{3}(t)$) \add{$\leq$ 0.523}, respectively.

\section{Discussion}

To our knowledge, this measurement is the first demonstration based on a simple qubit-fiber birefringence-reservoir model to study the non-Markovian effect on a two-photon polarization entangled state in which one photon from the pair is stored in a buffer made of different fiber lengths. We compare our model with the traditional atom-cavity-reservoir model.
The function  $\mathcal{P}_{a,sy}$ fits the experimental data better than \remove{the} $p_{3}(t)$. This is because the harmonic function of $\mathcal{P}_{a,sy}$ is a function of $\sqrt{t}$ due to the PMD effect in our model, while the harmonic function of $p_{3}(t) $ is a function of $t$ (see Methods for atom-cavity-reservoir  model). \remove{We first fit our experimental data with \remove{the} $\mathcal{P}_{a,sy}$, which is the summation of Eq.(3) and Eq.(4).} We find that the contribution from Eq.(4) is more dominant than \remove{the}  Eq.(3). This implies that the H and V polarization components of the buffered photon rotate in the same direction. This finding \remove{is a}\add{provides} new insight into the polarization properties of the buffered qubit \remove{that undergoes the}\add{in a} non-Markovian \remove{dynamics}\add{environment}. \remove{We then fit our experimental data with \remove{the} $p_{3}(t)$, which is the summation of Eq.(8) and Eq.(9). The obtained $\kappa$ is slightly above the threshold $\frac{\gamma_{\circ}}{4}$, which is the condition for an open quantum system to become non-Markovian.} Since our quantum system is operating close to the threshold $\kappa \sim \frac{\gamma_{\circ}}{4}$, the configuration of our experiment presents the possibility of naturally occurring crossover between the Markovian and non-Markovian processes.

\remove{We subtract the accidental coincidence \add{counts} in our data analysis so that the loss parameter $\mu$ is mainly contributed from the degradation of the polarization extinction ratio of the photonic qubit.
One can regard \remove{the} $\mu$ as the strength of the environment depolarization effect on the buffered photon. The obtained value of $\mu$ is below the induced attenuation value \add{of} 0.05$\,$dB/km caused by environmental factors such as temperature and humidity.} Our finding indicates that the density matrix $\rho(t)$ of the entangled state at the output of the buffer is related to its early \add{state} $\rho(t')$ in the buffer, where $t > t'$. This occurs when the environment depolarization effect of the fiber is more dominant than the PMD effect. We also demonstrate the Markovian scenario where the PMD effect did not exist by replacing the buffer with a free-space attenuator. \remove{representing a memory-less environment (fiber birefringence and fiber reservoir).}

\remove{We use the Werner state as the final state for the buffered idler photon and its paired signal photon.
The Werner state is entangled (inseparable) when the $P > \frac{1}{3}$. The concurrence is zero at $P = \frac{1}{3}$ which is the Werner's well-known separability criterion. The entanglement is related to the concurrence as $E_n(\rho_{AB})=H((1+\sqrt{1-\mathcal{C}_{n}^2})/2)$, where the $\mathcal{C}_{n}$ is the concurrence, and the function $H(x)=-x\log_2 x-(1-x)\log_2 (1-x)$ is the Shannon entropy. When the concurrence is zero, the entanglement is also zero at $P=\frac{1}{3}$. Based on this criterion, we can design the optimum fiber length of up to 190$\,$km for the buffer in which the state is still inseparable or entangled. This buffer length corresponds to the buffer time of 0.9$\,$ms.}

Several measures of detecting the non-Markovian dynamics have been proposed \cite{Luo_pra_2012,Plenio_PRL_2010,Piilo_16,Piilo_09,Piilo_10}. The most recent method uses quantum process capabilities of non-completely-positive (CP) processes \cite{Chan_24} for identifying the non-Markovian dynamics.
Our approach here \remove{of using}\add{uses} quantum state tomography to find the probability function $p(t)$, which leads us to the measures of quantum mutual information such as classical correlation and quantum discord.
\remove{Quantum discord is a better concept than entanglement to characterize the quantum correlation of an open quantum system \cite{Zurek_prl_2001}.}

\remove{We then use the quantum discord to characterize the quantum correlation of the entangled state in our buffer system.
The quantum discord is nonzero at $ P = \frac{1}{3}$ where the concurrence is zero.  (see Fig.S3 in the Supplementary). The non-zero quantum discord can surpass Werner's criterion on the optimum buffer time. The quantum discord implies that the quantum bi-partite correlation still exists for buffer time greater than 0.9$\,$ms.}

The possibility for the buffered photon to experience the case $\Delta\phi_{h} \neq \Delta\phi_{v}$ is rare for the single-mode fiber \remove{which is} used \remove{for}\add{in} communication networks. The reason is that the distributed fiber has a homogenous refractive index. The PMD effect of a light field in the single-mode fiber should not depend on the input polarization state of the light. However, the case $\Delta\phi_{h} \neq \Delta\phi_{v}$ may occur for the photon buffered in a loop of fiber more than one round trip. We are currently investigating the quantum buffer made of a fiber loop in a 2x2 switch configuration \cite{kflee24}, where the photon is buffered by passing through the same fiber loop multiple times.

\section{Conclusion}

\add{We have introduced a qubit-fiber-reservoir model to explore the non-Markovian dynamics of the two-photon polarization-entangled state. The non-Markovian probability function obtained from our approach fits well with the experimental data. One can design the model's parameters so that the crossover between Markovian and non-Markovian dynamics on the qubits can be controlled and utilized for modeling quantum networks \cite{Helsen_2023,Plenio_2014} and protecting quantum entanglement \cite{Wenjie_2023,Franco_23}.
Our analytical approach to deriving the probability function can be easily extended to two qubits interacting separately with two independent non-Markovian environments, with the idler and the signal buffered independently in two different buffers \cite{Breuer_oxford_2002,Petruccione_PRA_2006,Bellomo_PRL_2007,Bellomo_PRA_2008}. The non-Markovian effect will be strong, hence reducing the optimum buffer time.
Our approach provides new insight into the polarization properties of the buffered photon. This will be very beneficial for developing variable quantum buffers for controlling packet delays and allowing flexible contention management in quantum networks using classical-quantum data frames \cite{Shabani_22,BenYoo_2024} and a sequence of ordered signals \cite{Giovannetti_JPA_05}. Our model can be easily adapted into the formalism of tensor network states \cite{Lubasch_18} to model quantum processing.}

\section{Methods}

\textbf{Atom-Cavity-Reservoir Model}.
Our analytical approach is similar to a qubit of a two-level atom interacting in a high Q-cavity, which is coupled to the reservoir with the Lorentzian  spectral distribution \cite{Chen_PRA_14,Wang_scr_17}. The cavity-reservoir coupling gives rise to the effective spectral density in the form of, $(\gamma_{\circ}\Lambda^{2})/[(\omega_{\circ}-\omega)^{2}+\Lambda^{2}]$, where the parameter  $\gamma_{\circ}$ is the coupling constant between the cavity and the reservoir. The parameter $\Lambda$  is the spectral width of the reservoir. The atom transition frequency and the cavity are in resonance with $\omega_{\circ} = \omega_{\rm{cavity}} $. When $\Lambda$ becomes very large or approaches infinity, the reservoir is considered memoryless or Markovian. Then, the cavity is \remove{only} the main cause of the non-Markovian dynamics in the system. The diagonal (off-diagonal)density matrix elements of the atomic qubit \remove{are evolving}\add{evolve} in time $t$ as a function of $p(t) \propto e^{-\frac{\gamma_{\circ}t}{2}} [cos(\frac{\delta t}{4})+\frac{\gamma_{\circ}}{\delta}sin(\frac{\delta t}{4})]^{2}$ ($\sqrt{p(t)}$), respectively, where $\delta=\sqrt{16\kappa^{2}-\gamma_{\circ}^{2}}$, and $\kappa$ is the coupling constant between the atomic qubit and the cavity and \add{is} related to the decay rate of the excited state. The parameter $\delta$ can be used to determine whether the system is Markovian (  $4 \kappa < \gamma_{\circ}$) or non-Markovian ($4 \kappa  \geq \gamma_{\circ}$).

\textbf{The relationship of  $\gamma_{\circ}$ and $ \kappa$ on the Markovian and non-Markovian processes}.
The density matrix of the photon state is dependent on the probability function \cite{Chen_PRA_14,Wang_scr_17} $p(t) \propto e^{-\frac{\gamma_{\circ}t}{2}} [cos(\frac{\delta t}{4})+\frac{\gamma_{\circ}}{\delta}sin(\frac{\delta t}{4})]^{2}$, where $t$ is the propagation time of the photon in the fiber. The parameter $\delta=\sqrt{16\kappa^{2}-\gamma_{\circ}^{2}}$ can be used to determine whether our open quantum system is Markovian or non-Markovian.

For $ 4 \kappa < \gamma_{\circ}$ or $1 < \frac{\gamma_{\circ}}{4\kappa}$, $\delta$ \add{becomes imaginary}. The harmonic functions $sin(\cdot)$ and $cos(\cdot)$ become hyperbolic functions $cosh(\cdot)$ and $sinh(\cdot)$, respectively. Our quantum system is undergoing Markovian dynamics. This occurs when the PMD and the loss parameter are the dominant effects on the photonic qubit. Hence, the probability function $p(t)$ becomes an exponential function.
The polarization mode dispersion (PMD) effect that leads to the Markovian process, i.e., entanglement sudden death, has been experimentally observed \cite{Brodsky_OL_2011} and theoretically studied \cite{Brodsky_PRL_2011} on the fiber-based entangled photon pair. One can enhance the PMD effect on photon pairs by increasing their differential group delays, such as inserting a few polarization-maintaining fiber jumpers in their channels.

For $4 \kappa  \geq \gamma_{\circ}$, where the PMD effect is less dominant than the loss parameter, our system becomes non-Markovian. For simplicity, we assume $\gamma_{\circ} = 4\kappa/n$  where \add{the positive real number} $n > 1$, then $\delta = 4 \kappa \sqrt{1-\frac{1}{n^2}}$.
The probability function of the qubit is given by,

case 1: $n = \sqrt{2}$
\begin{eqnarray}
p_{1}(t) \propto e^{-\frac{\gamma_{\circ}}{2} t} [\rm{cos}(\frac{\kappa_1}{\sqrt{2}} t)+\rm{sin}(\frac{\kappa_1}{\sqrt{2}} t)]^{2}.
\label{eq:08}
\end{eqnarray}

case 2: large n
\begin{eqnarray}
p_{2}(t) \propto e^{-\frac{\gamma_{\circ}}{2} t} [\rm{cos}(\kappa_2 t)]^{2}.
\label{eq:09}
\end{eqnarray}
Then, the harmonic function of $p_{1,2}(t)$ depends only on the PMD effect.
However, our analytical approach can provide \remove{a} new physical insight into the coupling constant ($\kappa$) of the PMD and the loss ($\mu$) effects on the two-photon polarization entangled state when one photon from the pair is stored in a fiber-based quantum buffer.

\textbf{Amplitude Damping on one photon from the two-photon Werner State}.
The amplitude damping process on one qubit from two-qubit Werner state $\rho$ is given by,
\[
  \rho' =E_{\circ}\rho {E_{\circ}}^\dag +E_{1}\rho {E_{1}}^\dag =
  \begin{pmatrix}

    \frac{1+p}{4} & 0 & 0 & \frac{p\sqrt{1-\xi}}{2}\\
    0 & \frac{(1-p)(1-\xi)}{4}+\frac{(1+p)(\xi)}{4} & 0 & 0\\
    0 & 0 & \frac{1-p}{4} & 0\\
    \frac{p\sqrt{1-\xi}}{2} & 0 & 0 & \frac{(1+p)(1-\xi)}{4}+\frac{(1-p)(\xi)}{4}
 \end{pmatrix}
\]

where $E_{\circ}=(I\otimes\Gamma_{\circ})$ and $E_{1}=(I\otimes\Gamma_{1})$. The identity matrix $I=\begin{pmatrix}
    1 & 0 \\
    0 & 1
  \end{pmatrix}$. The amplitude damping operators are $\Gamma_{\circ} = \begin{pmatrix}
    1 & 0 \\
    0 & \sqrt{1-\xi}
  \end{pmatrix}$ and $\Gamma_{1} = \begin{pmatrix}
    0 & \sqrt{\xi} \\
    0 & 0
  \end{pmatrix}$, where the $\xi$ is the probability of depolarizing a photon.
We measure the density matrix $\rho'$ and obtain the density matrix elements: $\rho'_{11}=\frac{1+p'_{11}}{4}$, $\rho'_{22}=\frac{1-p'_{22}}{4}$, $\rho'_{33}=\frac{1-p'_{33}}{4}$, $\rho'_{44}=\frac{1+p'_{44}}{4}$, $\rho'_{14}=\frac{p'_{14}}{2}$, and  $\rho'_{41}=\frac{p'_{41}}{2}$. From each density matrix element, we write ${p'_{nm}}_{(n,m=1,2,3,4)}$ in terms of p of the Werner State $\rho$ as:

 \begin{eqnarray}
p'_{11} &=& p'_{33} = p\\
p'_{22} &=& p'_{44} = p (1-2\xi)\\
p'_{14} &=& p'_{41} = p\sqrt{1-\xi}\simeq p(1-\xi).
\label{eq:10}
\end{eqnarray}
where we approximate $p\sqrt{1-\xi}\simeq p(1-\xi)$ for $\xi < 0.04 (4\%)$. Our experimental data indicates that p drops less than $4\%$. We then take the average of the measured $\frac{p'_{11} + p'_{22} +p'_{33} + p'_{44} + p'_{14}+ p'_{41} }{6} = \frac{6p-5p\xi}{6} \sim p(1-\xi)=p(t)$. The $p(t)$ is the non-Markovian probability function. It shows that the p of the Werner state is changing by $(1-\xi)$.

\textbf{The probability function for  $\Delta\phi_{h} \neq \Delta\phi_{v}$ }.
For the situation $\Delta\phi_{h} \neq \Delta\phi_{v}$, Eq.(2) can be rewritten as
\begin{eqnarray}
\mathcal{P}_{\pm}^{f}
&=& e^{-2 \mu L} [ 2 + 2\, \rm{cos(\Delta\phi_v)} \rm{cos(\Delta\phi_h)} \nonumber\\
&-& 2\, \rm{sin(\pm \Delta\phi_h)} \rm{sin(\pm \Delta\phi_v)}- 2\, \rm{cos(\Delta\phi_h)} \rm{sin(\pm \Delta\phi_v)} \nonumber\\
&+& 2\, \rm{cos(\Delta\phi_h)} \rm{sin(\pm \Delta\phi_h)}- 2\, \rm{cos(\Delta\phi_v)} \rm{sin(\pm \Delta\phi_v)} \nonumber\\
&+& 2\, \rm{cos(\Delta\phi_v)} \rm{sin(\pm \Delta\phi_h)}].
\label{eq:11}
\end{eqnarray}

To simplify the above expression, we assume $\Delta\phi_{h(v)}$ is relatively small and substitute $\rm{cos(\Delta\phi)} = 1-\frac{\Delta\phi^2}{2}$ and $\rm{sin(\Delta\phi)}=\Delta\phi$ into the $\mathcal{P}_{\pm}^{f}$. The first few terms of the $\mathcal{P}_{\pm}^{f}$ are given as:

\begin{eqnarray}
\mathcal{P}_{\pm}^{f} \propto e^{-2 \mu L} [ 2 + \pm \mathcal{B} L^{\frac{1}{2}} \pm \mathcal{B'} L \mp \mathcal{C} L^{\frac{3}{2}} \pm \mathcal{C'} L^{2} \pm \mathcal{C} L^{\frac{5}{2}} \pm ..].
\label{eq:12}
\end{eqnarray}
This approximated probability function depends on the power of the fiber length $L^{\frac{m}{2}}$, where m is a positive integer.

\section{Data availability}
The data that supports the findings of this study are available from the corresponding author upon reasonable request.

\begin{acknowledgments}

\end{acknowledgments}

\newcommand{\noopsort}[1]{} \newcommand{\printfirst}[2]{#1}
   \newcommand{\singleletter}[1]{#1} \newcommand{\switchargs}[2]{#2#1}

\section{Acknowledgments}

This work was supported in part by the DOE (DE-SC0020537)

\section{Author Contributions}
K.F.L and P.K conceived the research.  K.F.L analyzed the data, wrote the paper and prepared the manuscript.

\section{Competing Interests}
The authors declare no competing interests.

\end{document}



\title{Supplementary Document: Non-Markovian Dynamics in Fiber Delay-Line Buffers}

\author{Kim Fook Lee and Prem Kumar}
\affiliation{%
Center for Photonic Communication and Computing, Department of Electrical Engineering and Computer Science, Northwestern University, 2145 Sheridan Road, Evanston, IL 60208-3112, USA}%
\email[]{kim.lee@northwestern.edu}

\date{Compiled \today}

\begin{abstract}
This document provides supplementary materials for "Non-Markovian Dynamics in Fiber Delay-Line Buffers".
\end{abstract}



\flushbottom

\maketitle

\thispagestyle{empty}

\clearpage
\setcounter{figure}{0} \renewcommand{\thefigure}{S\arabic{figure}}
\setcounter{table}{0} \renewcommand{\thetable}{S\Roman{table}}

\section*{Supplementary Information}
\textbf{Contents:}

%
%

\textbf{1. Fig. S1 The counter-propagating scheme (CPS) with the 300$\,$meters dispersion shifted fiber.}

\textbf{2. Fig. S2 The extrapolation plot of total, classical , quantum discord and concurrence with the buffer time $t$ up to 1.5$\,$ms.}

\textbf{3. Fig. S3 The extrapolation plot of total, classical , quantum discord, and concurrence as a function of probability.}

\newpage

\begin{figure}[ht]
\centering
\includegraphics[scale=0.7]{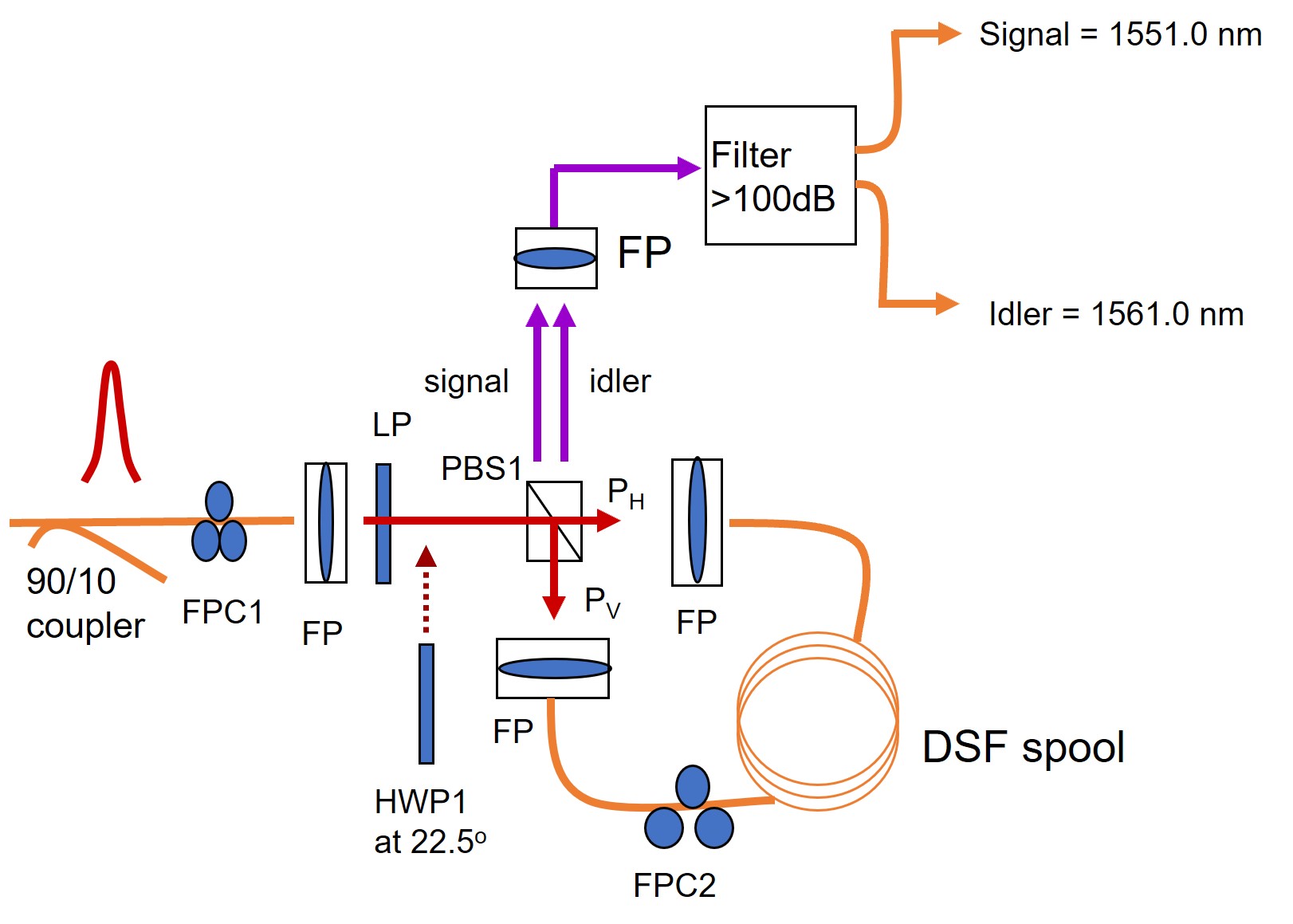}
\caption{\textbf{The counter-propagating scheme (CPS) with the 300$\,$m dispersion shifted fiber. The pump wavelength is at 1556$\,$nm. The pump laser is operated at 50$\,$MHz with the pulse duration of 5.0$\,$ps. The half-wave plate (HWP1) before the polarizing beam splitter (PBS1) is used to prepare the pump to horizontally($P_{H}$) and vertically ($P_{V}$) polarized pumps. The $P_{H}$ ($P_{V}$) pumps generated signal and idler in the clockwise (counter-clockwise) direction through the spontaneous four-wave mixing process in a 300$\,$m dispersion-shifted fiber. The two-photon polarization entangled state $|\psi_{\circ}\rangle=\frac{1}{\sqrt{2}}[|H_s H_i\rangle + |V_s V_i\rangle]$ is then created at the output of PBS1. FP: fiber port; FPC: fiber polarization controller.}}
\label{Figure1S}
\end{figure}

\newpage

\begin{figure}
\centering
\includegraphics[scale=0.5]{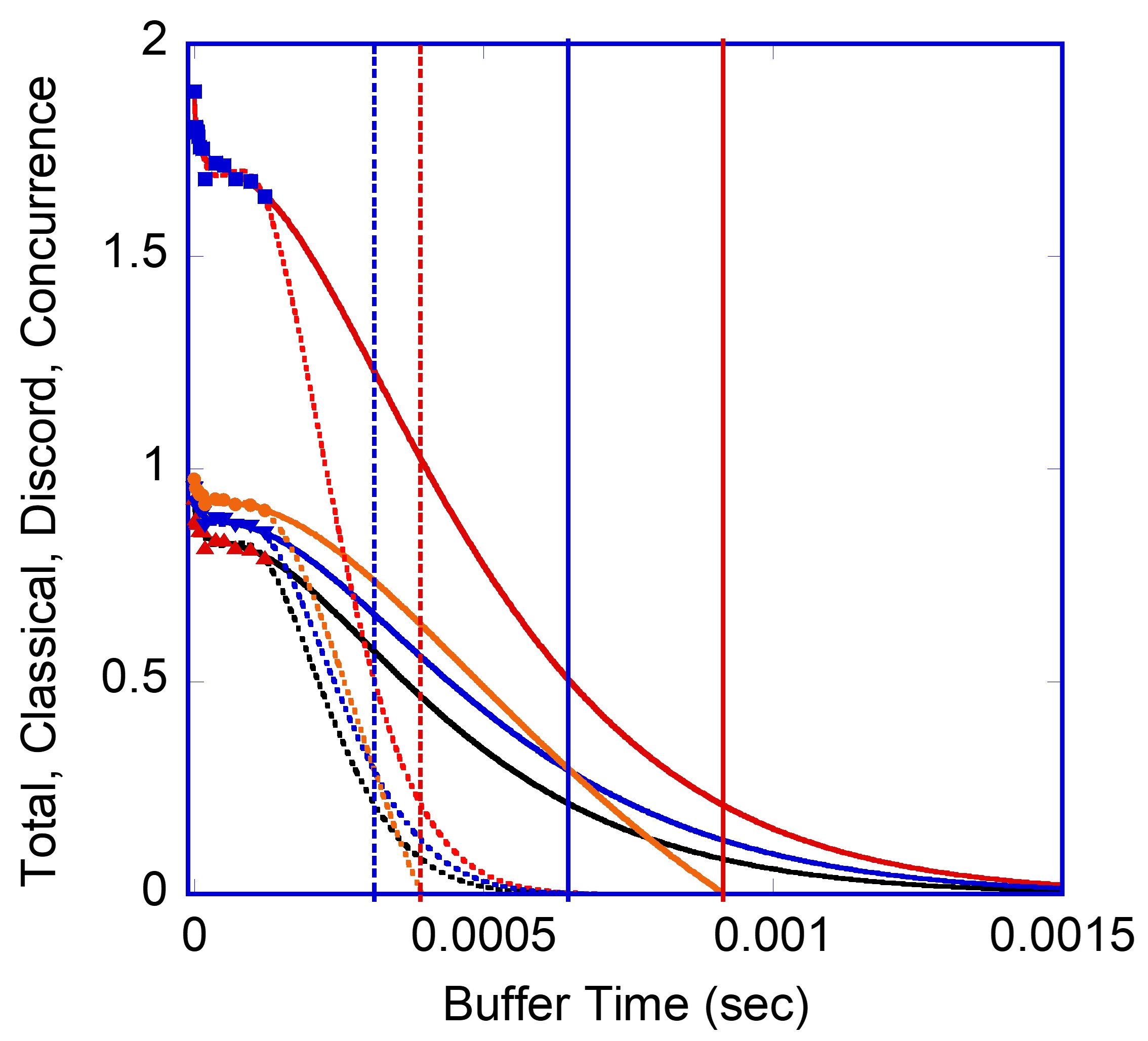}
\caption{\textbf{The extrapolation plot of total, classical , quantum discord and concurrence with the buffer time $t$ up to 1.5$\,$ms. The total $\mathcal{I}$ (red line), classical $\mathcal{C}$ (black line), quantum discord $\mathcal{Q}$ (blue line), and concurrence $\mathcal{C}_{n}$ (orange line) of the Werner state as a function of buffer time by using the equation of $\mathcal{P}_{a,sy}$. The experiment data: total (blue square), quantum discord (blue triangle), classical (red triangle), and concurrence (orange circle).
The vertical red line marks the location when  $\mathcal{P}_{a,sy}=0.333$ occurs at buffer time (fiber length) = 0.9$\,$ms (190$\,$km). The vertical blue line marks the location when $\mathcal{P}_{a,sy}=0.523$ occurs at buffer time (fiber length) = 0.7$\,$ms (145$\,$km). As for the extrapolation plot with the equation of $p_{3}(t)$, the total (red dotted line), quantum discord (blue dotted line), classical (black dotted line), and concurrence (orange dotted line). The vertically dotted red line marks the location when  $p_{3}(t)=0.333$ occurs at buffer time (fiber length) = 0.4$\,$ms (80$\,$km). The vertically dotted blue line marks the location when $p_{3}(t)=0.523$ occurs at buffer time (fiber length) = 0.3$\,$ms (64$\,$km).}}
\label{Figure2S}
\end{figure}

\newpage

\begin{figure}[ht]
\centering
\includegraphics[scale=0.5]{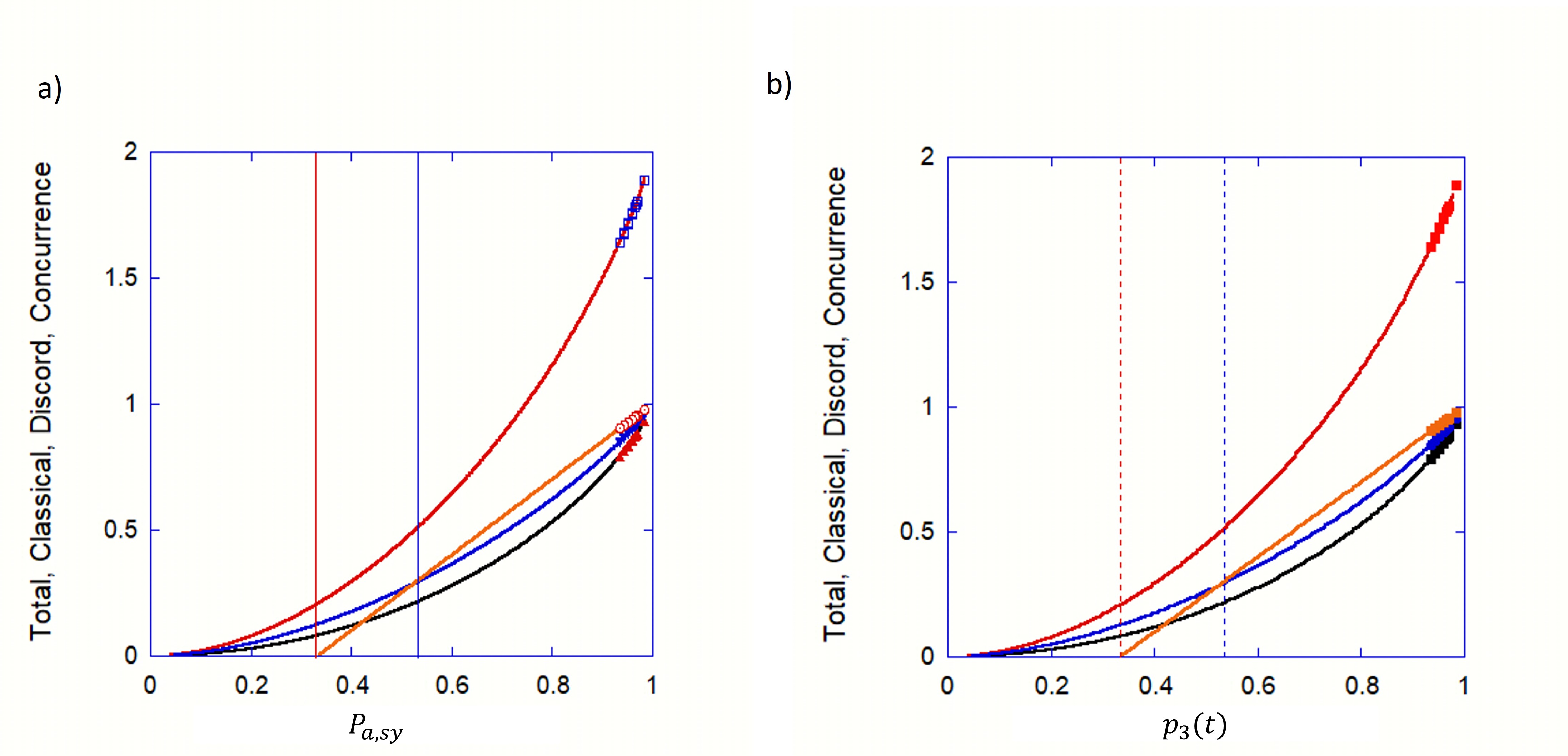}
\caption{\textbf{(a)The extrapolation plot of total (red line), classical (black line) , quantum discord (blue line), and concurrence (orange line) as a function of probability $\mathcal{P}_{a,sy}$. The vertical red line marks the location when  $\mathcal{P}_{a,sy}=0.333$ occurs at buffer time (fiber length) = 0.9$\,$ms (190$\,$km). The vertical blue line marks the location when $\mathcal{P}_{a,sy}=0.523$ occurs at buffer time (fiber length) = 0.7$\,$ms (145$\,$km). (b) As for the extrapolation plot with the equation of $p_{3}(t)$, the vertical red dotted line marks the location when  $p_{3}(t)=0.333$ occurs at buffer time (fiber length) = 0.4$\,$ms (80$\,$km). The vertical blue dotted line marks the location when $p_{3}(t)=0.523$ occurs at buffer time (fiber length) = 0.3$\,$ms (64$\,$km).}}
\label{Figure3S}
\end{figure}

\newpage

\newpage